\documentclass[12pt]{article}
\usepackage{amsfonts}
\usepackage{amssymb}

\usepackage{epsf}

\begin{document}
\thispagestyle{empty}

\noindent{\large\bf Instability of a Landau Fermi liquid\\
 as the Mott insulator is approached}

\vspace{2cm}

\noindent {\bf N. Furukawa$^*$ \ and \ T.M.Rice}

\vspace{.7cm}

\noindent {\bf Theoretische Physik, ETH-H\"onggerberg, \\
CH-8093   Z\"urich, Switzerland}

\vspace{7cm}

\noindent {\bf Abstract} \ \ We examine a two-dimensional Fermi liquid
with a Fermi surface which touches the Umklapp surface first at the 4
points $(\pm \pi/2, \ \pm \pi/2)$ as the electron density is
increased. Umklapp processes at the 4 patches near $(\pm \pi/2,\  \pm
\pi/2)$ lead the renormalization group equations to scale to strong
coupling resembling the behavior of a 2-leg ladder at
half-filling. The incompressible character of the fixed point causes a
breakdown of Landau theory at these patches. A further increase in
density spreads the incompressible regions so that the open Fermi
surface shrinks to 4 disconnected segments. This non-Landau state, in which
parts of the Fermi surface are truncated to form an insulating spin
liquid, has many features in common with phenomenological models recently
proposed for the cuprate superconductors. 

\vspace{2.5cm}

\hrule

\bigskip

\noindent
$^*$ Permanent Address:\\
$\phantom{m}$Institute for Solid State Physics, Univ.~of Tokyo,
Minato-ku, Tokyo 106-8666, Japan. 

\vfill\eject

\baselineskip=.8cm

\noindent
One of the key issues in high-T$_{\rm c}$ superconductivity is the
nature of the anomalous normal state which shows many indices of
non-Landau-Fermi liquid behavior. This contrasts with the behavior of
many other transition metal oxides which have strongly renormalized
Landau-Fermi liquids near to the Mott transition but no
superconductivity [1]. Haldane [2] has shown that Landau theory is generally
valid in two dimensions. This has led to a search for some instability
of the Fermi liquid with increasing electron density in the overdoped
cuprates. In this letter we show how Umklapp scattering can cause an
instability of the Fermi surface of a 2-dimensional metal as the Mott
state is approached. This instability need not involve long range
magnetic order but it causes a charge gap and truncation of parts of
the Fermi surface. This type of behavior has been documented in a
lightly doped 3-leg ladder where the two even parity channels form an
insulating spin liquid (ISL) leaving a Fermi surface only in the odd parity
channel [3, 4]. Here we discuss how a similar behavior can arise in two
dimensions.

We start with a general 2-dim.~dispersion relation, for
example a form, $\varepsilon({\bf k}) = - 2 t (\cos k_x + \cos k_y) -
4 t' \cos k_x \cos k_y$ with $t(t')$ as (next) nearest neighbor
hopping matrix elements. 
Taking $t>0$ and $t'>0$ and increasing the electron density, $n$, leads
to a Fermi surface which touches the surface at which Umklapp
processes are allowed first at the 4 points $(\pm \pi/2, \ \pm
\pi/2)$. Following Haldane [2] we divide the Fermi surface into patches and
examine the patches near the 4 points $(\pm \pi/2, \ \pm \pi/2)$.
These 4 patches on the Fermi surface are connected through Umklapp
processes which leads us to examine the renormalization group (RG)
equations for the coupling constants. The RG equations have
similarities to those for a 2-leg ladder at half-filling which also
has 4 Fermi surface points and which are known to scale to a strong
coupling solution.

The 4 patches around $(\pm \pi/2, \ \pm \pi/2)$ are sketched in
Fig.~1. The size of a patch is defined by a wavevector cutoff, $k_c$,
and within each patch $\alpha (\alpha=1,\ldots,4)$ the electron energy
relative to the chemical potential, $\mu$ is expanded as 
\begin{equation}
\varepsilon_\alpha({\bf q}) - \mu = v q_\alpha + u q_{\bot,\alpha}^2
\end{equation} 
where ${\bf q}$ is the wavevector measured from the center of
$\alpha$-th patch, and $q_\alpha \ (q_{\bot,\alpha})$ is the component
of ${\bf q}$ normal (tangent) to the Fermi surface at the center of
the patch. The Fermi velocity is given by $v$, and the energy cutoff
is $E_0=vk_c$. We define $v^*=\pi v/k_c$, and hereafter take $\pi
v^*=1$ as the unit of energy. 

The linear dispersion relation leads to logarithmic anomalies in the
particle-hole (Peierls) and particle-particle (Cooper) channels as in
one dimension but the transverse dispersion introduces an infrared
cutoff, $E_T$, with a magnitude $E_T \approx u k_c^2$. The
non-interacting susceptibility in the Peierls channel takes the form
$\chi_p (\omega) = 1/2 \ln ({\rm max} (\omega, E_T)/E_c)$. In the
parameter region $\omega > E_T$, the infrared cutoff from the
transverse dispersion can be ignored and a set of RG equations can be
derived as in one dimension [5].

In Fig.~2 we define the normal vertices
$g_{1}$, $g_{2}$ and $g_{1\rm r}$ as well as
Umklapp vertices $g_3$, $g_{3\rm p}$ and $g_{3\rm x}$.
Other interactions are not treated here since they are 
irrelevant within the framework of a one-loop approximation.
Summing up all one-loop diagrams, we obtain the RG equations
\begin{eqnarray}
 \dot{g}_1 &=& 
   g_1{}^2 + g_{1\rm r}{}^2 + 2 g_{3\rm x}{}^2 -2 g_{3\rm x}g_{3\rm p},\\
 \dot{g}_2 &=& 
   \frac12(
     g_1{}^2 + 2 g_{1\rm r}{}^2 - g_3^2 - 2 g_{3\rm p}{}^2
   ),\\
 \dot{g}_{1\rm r} &=&
   (g_1+g_2) g_{1\rm r}  ,\\
 \dot{g}_3 &=& 
   (g_1 - 2g_2) g_3 
   + 2 g_{3\rm x}{}^2 - 2 g_{3\rm x}g_{3\rm p} - g_{3\rm p}{}^2,\\
 \dot{g}_{3 \rm x} &=& 
   2 g_1 g_{3\rm x} - g_1 g_{3\rm p} - g_2 g_{3\rm x}
   + g_3 g_{3\rm x} - g_3 g_{3\rm p},\\
 \dot{g}_{3 \rm p} &=& 
   - (g_2+g_3) g_{3\rm p}.
\end{eqnarray}
Here $\dot{g}_i \equiv x ({\rm d}g_i)/({\rm d}x)$ and $x=\omega/E_0$.

Note these equations differ from those obtained by Zheleznyak {\em et al.}
[6] who earlier considered a model with 4 flat 
Fermi surface patches. In their model the patches were oriented perpendicular
to the (1,0) and (0,1) directions and as a consequence Umklapp
processes connecting perpendicular patches did not enter the
RG equations which are then more closely related to those for a
single chain at half-filling. Eqns. (2-7) coincide with those
derived previously by Houghton and Marston [7]
who considered the problem of a lightly doped flux state
rather than the present limit of a heavily doped model with
next-nearest-neighbor hopping.

We take repulsive
 Umklapp interactions, as $g_3 = g_{3\rm x} = g_{3 \rm p} = U$,
and treat $g_1$, $g_2$ and $g_{1\rm r}$ as parameters.
The fixed points are obtained by numerically integrating 
the RG equations.
In a substantially wide region around $g_1 \sim g_2 \sim g_{1\rm r} \sim U$, 
we find a strong coupling fixed point where
both normal and Umklapp vertices diverge.
The corresponding phase diagram is shown in Fig.~3.
The asymptotic behavior of the vertices is given by
$g_i = g_{i}^0 \Lambda /[1 +\Lambda \log(\omega/E_0)]$,
where
$(g_{1}^0,g_{2}^0,g_{1\rm r}^0, g_3^0,g_{3\rm p}^0, g_{3\rm x}^0)
 = (\frac{1}{14},\frac{5}{14},0,\frac{9}{14},\frac17\sqrt{\frac{15}{2}}
,\frac{1}{14}\sqrt{\frac{15}{2}})$.
A singularity appears at 
$\omega \sim \omega_c = E_0\exp(-1/\Lambda) $
where $\Lambda \propto U$. 
In two dimensions, such an anomaly
at finite $\omega$ is an artifact of the one-loop calculation
and  higher order terms will shift it to $\omega=0$. Nevertheless, $\omega_c$
represents the energy scale where the system  crosses over 
from weak coupling to strong coupling.
We will explicitly assume that the interactions $\sim U$ are strong enough so
that $\omega_c > E_T$ in which case the existence of a finite curvature
becomes irrelevant at the strong coupling fixed point. In contrast, if
the system had scaled to weak coupling, then $E_T$ would always
remain relevant. There is a limit with weak interactions and a
dispersion relation with $t' \ll t$ where both conditions, $E_T (=
2t' k_c^2) \ll \omega_c$ and $U \ll E_0$, are satisfied and our approach
based on one-loop RG equations is justified. 
We speculate that the qualitative nature of the anomaly
obtained in this weak coupling region is also present
 in the strong coupling region $U \gtrsim t,t'$.

We now discuss the nature of the fixed point 
through  the anomalies in the susceptibilities.
Due to the nesting behavior in the non-interacting case,
there may exist anomalies in the spin susceptibility 
($\chi_{\rm s}$) and the charge susceptibility ($\chi_{\rm c}$)
at $q =(\pi,\pi)$. Within the one-loop calculation, we find
\begin{eqnarray}
 \chi_{\rm s}(\pi,\pi) &\propto& (\omega-\omega_{\rm c})^{\alpha_{\rm s}},
   \quad 
   \alpha_{\rm s} = {-g_2^0-g_3^0-2 g_{3\rm p}^0},\\
 \chi_{\rm c}(\pi,\pi)&\propto&  (\omega-\omega_{\rm c})^{\alpha_{\rm c}},
   \quad 
   \alpha_{\rm c} = {2g_1^0-g_2^0+g_3^0-2g_{3\rm p}^0+4g_{3\rm x}^0}.
\end{eqnarray}
The superconducting susceptibilities
for $s$-, $p$- and $d$-wave pairing
($\Delta_s$, $\Delta_p$ and $\Delta_d$, respectively)
also behave as
\begin{eqnarray}
 \Delta_{s} &\propto& (\omega-\omega_{\rm c})^{\alpha_{\rm ss}},
   \quad 
   \alpha_{\rm ss} = {g_{2}^0+g_{1}^0+2g_{1{\rm r}}^0}, \\
 \Delta_{p} &\propto& (\omega-\omega_{\rm c})^{\alpha_{\rm ps}},
   \quad 
   \alpha_{\rm ps} = {g_2^0-g_1^0}, \\
 \Delta_{d} &\propto& (\omega-\omega_{\rm c})^{\alpha_{\rm ds}},
   \quad 
   \alpha_{\rm ds} = {g_2^0+g_1^0-2g_{1\rm r}^0}.
\end{eqnarray}
At the fixed point with strong Umklapp coupling described above,
the leading divergence is observed in the spin susceptibility with 
the exponent $\alpha_{\rm s}=-1.782$, while the exponents for
charge and superconducting susceptibilities are positive so that
these susceptibilities do not diverge at the critical point.

The uniform spin $(\chi_s(0))$ and charge $(\kappa)$ [8, 9]
susceptibilities are  also of interest. 
As in the case of 1d chain system, we have
spin gap behavior when there is a divergence in  $g_1$ 
and charge gap behavior from $g_3$, 
\begin{eqnarray}
 \chi_{\rm s}(0) &\propto& (\omega-\omega_{\rm c})^{({g_1^0})^2/2},
  \label{EqUniformSpin}\\
 \kappa&\propto& (\omega-\omega_{\rm c})^{(g_{3}^0){}^2/4}.
\end{eqnarray}
Here we approach $\omega_{\rm c}$ by decreasing both $\omega$ and $q$
with $q/\omega$ being fixed.
In the present case, both $g_1$ and $g_3$ flow to strong coupling
which indicates a tendency to open up both spin and charge gaps.

We now compare the present results to those of a two-leg ladder at
half-filling. In this case, as Balents and Fisher have shown [10], there are
9 vertices which are relevant within a one-loop calculation. Again the
flow is to strong coupling in backward and Umklapp scattering
channels. In this case the properties of the strong coupling fixed
point are well established. The system is an insulating spin liquid
(ISL) with both spin and charge gaps (C0S0 in the Balents-Fisher
notation) and is an example of a short range RVB (Resonant Valence
Bond) state first proposed by Anderson for a $S=1/2$ Heisenberg
model [11]. The spin susceptibility $\chi_s (\pi,\pi)$ is strongly enhanced
but remains finite. 

In the present case we cannot be sure of the spin
properties from the one-loop calculations especially since the spin
susceptibility at $(\pi,\pi)$ and $(0,0)$ behave in a contradictory
fashion. What is certain is the scaling to strong coupling with
diverging Umklapp scattering. This gives us confidence in the result
that the compressibility $\kappa = d k_{F,\alpha} / d\mu \to 0$ at
the  fixed point as it does in the two-leg ladder at
half-filling. This has several 
profound consequences. First the condensate that forms is pinned and
insulating. Secondly when additional electrons are added to the system
the Fermi surface does not simply expand along the $(\pm 1, \pm 1)$
directions beyond the $(\pm \pi/2, \ \pm \pi/2)$ points as would
happen for noninteracting electrons. Instead the charge gap and
vanishing $dk_{F\alpha}/d\mu$ force the additional electrons to be
accommodated in the rest of Fermi surface.

Our proposal that the
strong coupling fixed point is in the same universality class (C0S0)
as that of the two-leg ladder at half-filling differs from 
 the conclusions of Zheleznyak {\em et al.} [6]
who found antiferromagnetic order in their model
 which in turn is rather related to
the class of fixed points of the single chain at half-filling (C0S1).
Note that the finite curvature in the present model, which kills
nesting properties, acts to suppress the Peierls channel away
from the 4 patches and therefore the tendency to
antiferromagnetic order. A similar conclusion was reached in ref.~[6]
when a finite curvature was included.

To examine what happens next we increase the electron density such that
there are 8 points on the Umklapp surface which intersect the
non-interacting Fermi surface at a finite angle (see Fig.~4a). This
leads us to examine an 8-patch model allowing for Umklapp scattering
processes. Within a one-loop calculation, the 8-patch model has 9
relevant coupling constants, defined in Fig.~5, and their RG equations
are given by 
\begin{eqnarray}
 \dot{g}_1 &=& 
  g_1{}^2 + g_{1\rm x} g_{2\rm x} + g_{1\rm s}g_{1\rm l} + g_{1\rm r}{}^2 
  + g_{3\rm x}{}^2 - g_{3\rm p}g_{3\rm x},\\
 \dot{g}_2 &=& 
   \frac12 ( g_{1}{}^2
         + g_{1\rm x}{}^2 + g_{2 \rm x}{}^2 + g_{1\rm s}{}^2 + g_{1\rm l}{}^2
         + 2 g_{1\rm r}{}^2 
         - g_{3\rm p}{}^2
      ) ,\\
 \dot{g}_{1\rm x} &=&
   g_1 g_{2\rm x} + g_2 g_{1 \rm x} 
   + (g_{1 \rm s} +g_{1 \rm l}) g_{1 \rm r} ,\\
 \dot{g}_{2\rm x} &=&
   g_1 g_{1\rm x} + g_2 g_{2 \rm x} 
   + (g_{1 \rm s} +g_{1 \rm l}) g_{1 \rm r}, \\
 \dot{g}_{1\rm s} &=&
   g_1 g_{1 \rm l} + g_2 g_{2 \rm s} 
   + (g_{1 \rm x} + g_{2 \rm x}) g_{1 \rm r},\\
 \dot{g}_{1\rm l} &=&
   g_1 g_{1 \rm s} + g_2 g_{2 \rm l} 
   + (g_{1 \rm x}+ g_{2 \rm x}) g_{1 \rm r}, \\
 \dot{g}_{1\rm r} &=&
   2 (g_1 + g_2) g_{1\rm r} 
   + (g_{1\rm x } + g_{2 \rm x}) (g_{1 \rm s}+ g_{1 \rm l}) ,\\
 \dot{g}_{3\rm p} &=& 
   - g_2 g_{3\rm p} ,\\
 \dot{g}_{3\rm x} &=&
  (2 g_1 -g_2) g_{3 \rm x} - g_1 g_{3 \rm p}.
\end{eqnarray}

For Hubbard-like interactions, $g_i = U$, 
numerical integration shows that the interactions
flow to a strong coupling fixed point with divergent $g_1$, $g_2$,
$g_{3\rm x}$ and $g_{3 \rm p}$ with prefactors as
$g_1^0 =\frac13(-2+\sqrt{10})$, $g_2^0 = 1$, 
and $g_{3\rm p}^0= 2 g_{3\rm x}^0= \frac23\sqrt{8-\sqrt{10}}$,
while the other couplings flow to zero.
Examining the susceptibilities gives us again the result
that the spin susceptibility diverges most strongly with an exponent
$\alpha_s = -g_2^0=-1$ . However the wavevector is no longer $(\pi,\pi)$
but the incommensurate wavevector connecting patches that span the
Fermi surface. The uniform spin susceptibility $\chi_s(0)$ diverges to
zero but the compressibility $\kappa$ is not renormalized to this
order. 

The scaling of the one-loop equations to a strong coupling fixed point,
even when we start with a non-interacting Fermi surface, makes it
necessary to consider the strong coupling behavior further. The
restriction that we found in the 4-patch model, which prevents the
Fermi wavevector along $(\pm 1, \ \pm 1)$ directions from extending
past the Umklapp surface, strongly suggests that this behavior will
spread out laterally as indicated in Fig.~4b. Again we can draw a
parallel to the slightly doped 3-leg ladder where in strong coupling a
C1S1 phase containing an ISL with commensurate filling in the even
parity channels remains stable up to a critical density. This contrasts with
the one-loop result which gives a C2S1 phase with holes immediately
entering both odd and even-parity channels. Actually we can examine
the strong coupling limit self-consistently. If we start from a
dispersion of the form Fig.~4b in a 8-patch model, then we find that
$\kappa \to 0$. So a lateral spreading of the truncated region of the
Fermi surface, as shown in Fig.~4b, is self-consistent. 

While a one-loop calculation is of limited validity at a strong
coupling fixed point, there are a number of conclusions that can be
drawn. First the breakdown of Landau theory, as the system scales to a
fixed point with strong Umklapp scattering is certain. Less certain is
 the question
whether it has long range spin order or not. Even if it has, it is not
the essence of the fixed point, which is the divergence of
Umklapp scattering as the fixed point is approached.
The most likely form remains an ISL spreading
across the Fermi surface which successively truncates the Fermi
surface as shown schematically in Fig.~4b. This is consistent with the
result that $\chi_s (0) \to 0$ 
always. The condensate is pinned and insulating due to the
Umklapp scattering. We note that an ISL, which truncates part of the Fermi
surface, is not characterized by any simple broken symmetry or order
parameter. It is not amenable to a simple mean field or Hartree-Fock
theory since there are no anomalous averages. Also the onset of the
ISL  is a
crossover rather than a phase transition at finite temperature. Lastly
we should remark that the ISL is not incompressible since it can
expand or contract laterally by exchanging electrons with the open
parts of the Fermi surface. 

The Fermi surface in the cuprates has a different form with $t'<0$ and
not $t'>0$ as we investigated. In this case the Fermi surface first
touches the Umklapp surface at the saddle points $(\pi,0)$ and
$(0,\pi)$. This leads to complications in the analysis due to the
presence of $\ln^2 (\omega/E_0)$ terms in the RG equations which will
require additional analysis in the future. 
Dzyaloshinskii [12] has made a detailed analysis
of a weak coupling fixed point for this model leading to a form of
Luttinger liquid behavior. Our result for the 8-patch model lead us to
consider rather possible strong coupling fixed points with divergent
Umklapp scattering which would introduce a charge gap at the 
Fermi surface.
The open Fermi surface will then be 4 disconnected segments centered
on $(\pm \pi/2, \ \pm \pi/2)$ as sketched in Fig.~4c, which is
similar to the results of a recent SU(2) gauge theory calculation [13]. 
Signs of such behavior are also evident in a recent analysis of the
momentum distribution using a high temperature series by
Putikka {\em et al.}  [14].
Note they include only nearest-neighbor terms in the kinetic energy
({\em i.e.} $t'=0$) which is a special limit from the present
point of view.

If we accept the premise that in the cuprates Umklapp scattering
stabilizes an ISL in the vicinity of the saddle points, then there are
some interesting consequences. For example, the ISL provides a
microscopic justification for some phenomenological
models. Geshkenbein, Ioffe and Larkin [15] proposed a model of preformed
pairs in the vicinity of the saddle point but with a very large mass
to suppress their contribution to the conductivity in the normal
phase. Similarly the ISL can provide a reservoir of pairs that act to
induce superconductivity in the open parts of the Fermi surface but
with the difference that here electron pairs cannot be scattered into
the ISL, only hole pairs. In fact experience with the ISL in the 2-leg
ladder shows that it is preferable energetically to add holes in pairs
to an ISL. Such processes will then by an efficient mechanism cause
pairing on the open segments of the Fermi surface. Also recently
Murakami and Fukuyama [16] included Umklapp scattering in a mean 
field treatment and found that it enhanced $d_{x^2-y^2}$-pairing.  In
the normal state 
there is a close similarity to a phenomenological model proposed by
Ioffe and Millis [17], to explain the anomalous transport properties. Here
also the Fermi surface segments have usual quasi-particle properties
(i.e. there is no spin-charge separation) and the scattering rate will
vary strongly since Umklapp processes will lead to strong scattering
at the end of the segments where they meet the Umklapp surface. These are
key features of the phenomenological Ioffe-Millis model.
Ioffe and Millis justified their model by a comparison to the
tunneling and ARPES experiments [18, 19] which show a single particle gap
opening in the vicinity of the saddle points similar to the form in Fig.~4c.
Lastly we refer the reader to the very recent preprint by
Balents, Fisher and Nayak [20]
which introduces the concept of a Nodal liquid 
with properties similar to the ISL discussed above.

In conclusion we have shown that when the Fermi surface approaches 
the Umklapp surface, the addition of Umklapp scattering can cause a
breakdown of Landau theory. The Fermi surface is truncated by the
formation of a pinned and insulating condensate. We have
given arguments that the spin properties are those of an insulating
spin liquid. This microscopic model has a lot in common with some
recent phenomenological models so that we believe it can form the
basis for a theory of the cuprates. 

\vspace{.5cm}

\noindent
{\bf Acknowledgments}\\
We wish to thank S. Haas, D. Khveshchenko, M. Sigrist and E. Trubowitz for
stimulating conversations. \\
N.F. is supported by a Monbusho Grant for overseas research.

\newpage

\noindent
{\bf References}
\begin{description}
\item[[1]] \ M. Imada, A. Fujimori and Y. Tokura, Rev. Mod. Phys, in press.

\item[[2]] \ F.D.M. Haldane, in 
Proc. of the International School of Physics ``Enrico Fermi'',
Course 121, 1992, ed. J.R. Schrieffer and R.A. Broglia
(North Holland, New York, 1994).  See also
 J. Fr\"ohlich and  R. G\"otschmann, Phys. Rev. B{\bf 55}, 6788 (1997).

\item[[3]] \ T.M. Rice, S. Haas, M. Sigrist and F.C. Zhang,
Phys. Rev. B{\bf 56} 14655 (1997).

\item[[4]] \ S.R. White and D.J. Scalapino,
Phys. Rev. B{\bf 57}, 3031 (1998).

\item[[5]] \ J. S\'olyom,
Adv. in Phys. {\bf 28}, 201 (1979).

\item[[6]] \ A. T. Zheleznyak, V. M. Yakovenko and I. E. Dzyaloshinskii,
Phys. Rev. B{\bf 55}, 3200 (1997).

\item[[7]] \ A. Houghton and J. B. Marston, Phys. Rev. B48, 7790 (1993).

\item[[8]] \ H. Fukuyama, T.M. Rice, C.M. Varma and B.I. Halperin,
Phys. Rev. B{\bf10}, 3775 (1974).

\item[[9]] \ M. Kimura, Prog. Theo. Phys. {\bf 53}, 955 (1975).

\item[[10]] L. Balents and M.P.A. Fisher,
Phys. Rev. B{\bf 53}, 12133 (1996); H.-H. Lin, L. Balents and M.P.A. Fisher,
Phys. Rev. B{\bf 56}, 6569 (1997).

\item[[11]] P.W. Anderson, Science {\bf 235}, 1196 (1987).

\item[[12]] I. Dzyaloshinskii, J. Phys. I France {\bf 6}, 119 (1996).

\item[[13]] P. A. Lee and X. G. Wen, Phys. Rev. Lett.
{\bf 78}, 4111 (1997).

\item[[14]] W. O. Putikka, M. U. Luchini and R. R. P. Singh,
preprint, cond-mat/9803141.

\item[[15]] V.G. Geshkenbein, L.B. Ioffe and A.I. Larkin, 
Phys. Rev. B{\bf 55}, 3173 (1997).

\item[[16]]
M. Murakami and H. Fukuyama, preprint,  cond-mat/9802009.

\item[[17]] L.B. Ioffe and A.J. Millis, preprint, cond-mat/9801092.

\item[[18]] Ch.~Renner, B. Revaz, J.-Y. Genoud, K. Kadowski and
{\O}. Fischer, Phys. Rev. Lett. {\bf 80}, 149 (1998).

\item[[19]] A.G. Loeser, Z.-X. Shen, D.S. Dessau, D.S. Marshall,
C.H. Park, P. Fournier and A. Kapitulnik, Science, {\bf 273}, 325 (1996); 
H. Ding, T. Yokoya, J.C. Campuzano, T. Takahashi, M. Randeria, M.R. Norman,
T. Mochiku, K. Kadowaki and J. Giapintzakis,
Nature, {\bf 382}, 51 (1996).

\item[[20]] 
L. Balents, M. P. A. Fisher and  C. Nayak,
preprint, cond-mat/9803086.

\end{description}

\pagebreak
\noindent

\pagebreak
\epsfxsize=7.5cm
\epsfbox{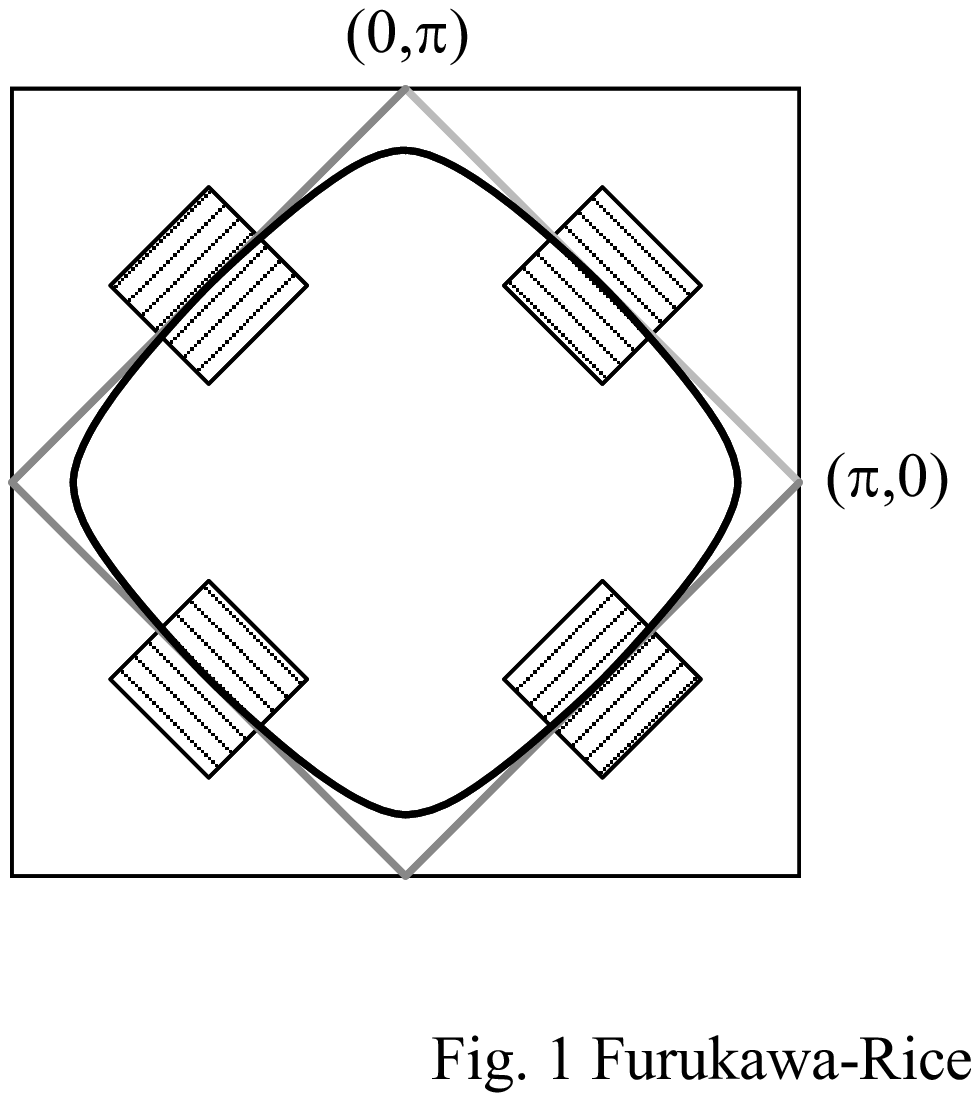}

Definitions of 4 patches, shown as hatched rectangular areas.
The bold curve represents the 2-dimensional Fermi surface which touches
the points $(\pm \pi/2, \pm\pi/2)$. The grey lines show the
Umklapp surface where Umklapp processes are allowed.

\newpage

\epsfxsize=12cm
\epsfbox{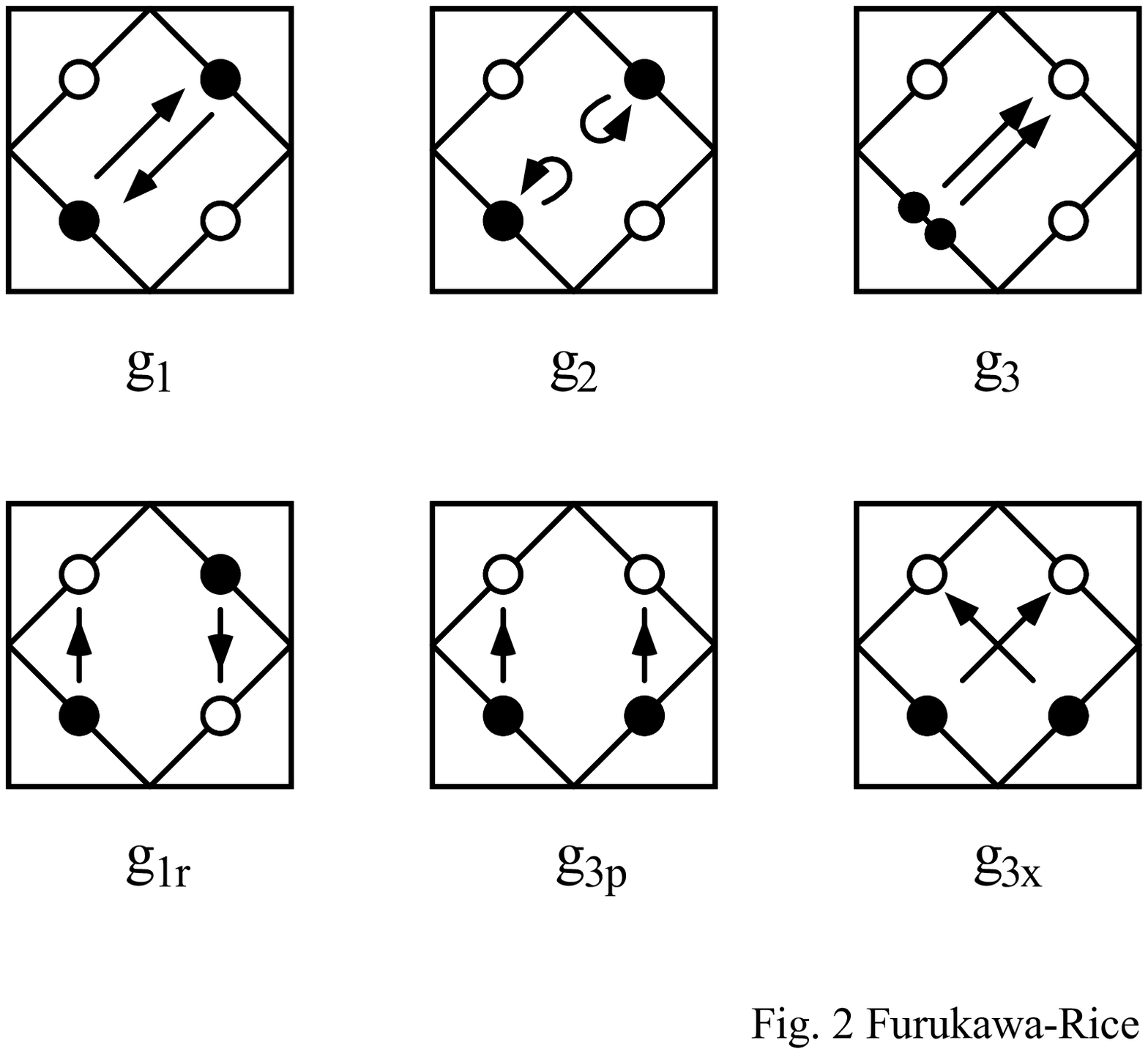}

The definitions of vertices for the 4-patch model.

\newpage
\epsfxsize=9cm
\epsfbox{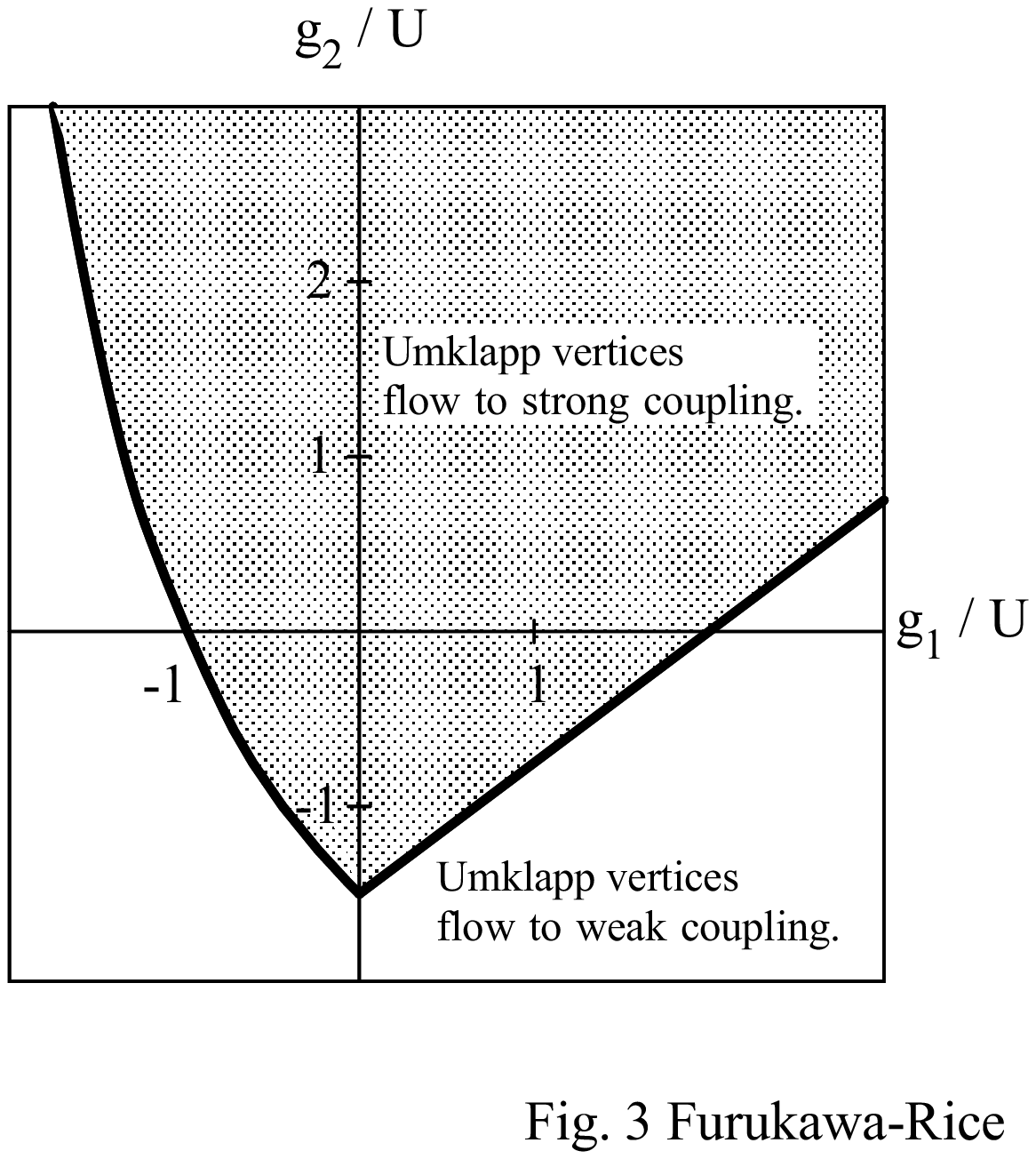}

Phase diagram of the 4-patch model at
$g_3 = g_{3\rm x} = g_{3\rm p} = U$.
The hatched area shows the region where Umklapp interactions flow to strong
coupling. Here, we take $g_{1\rm r} = g_1$, but the hatched region
does not change qualitatively if we take other values for $g_{1\rm r}$.

\newpage
\epsfxsize=15cm
\epsfbox{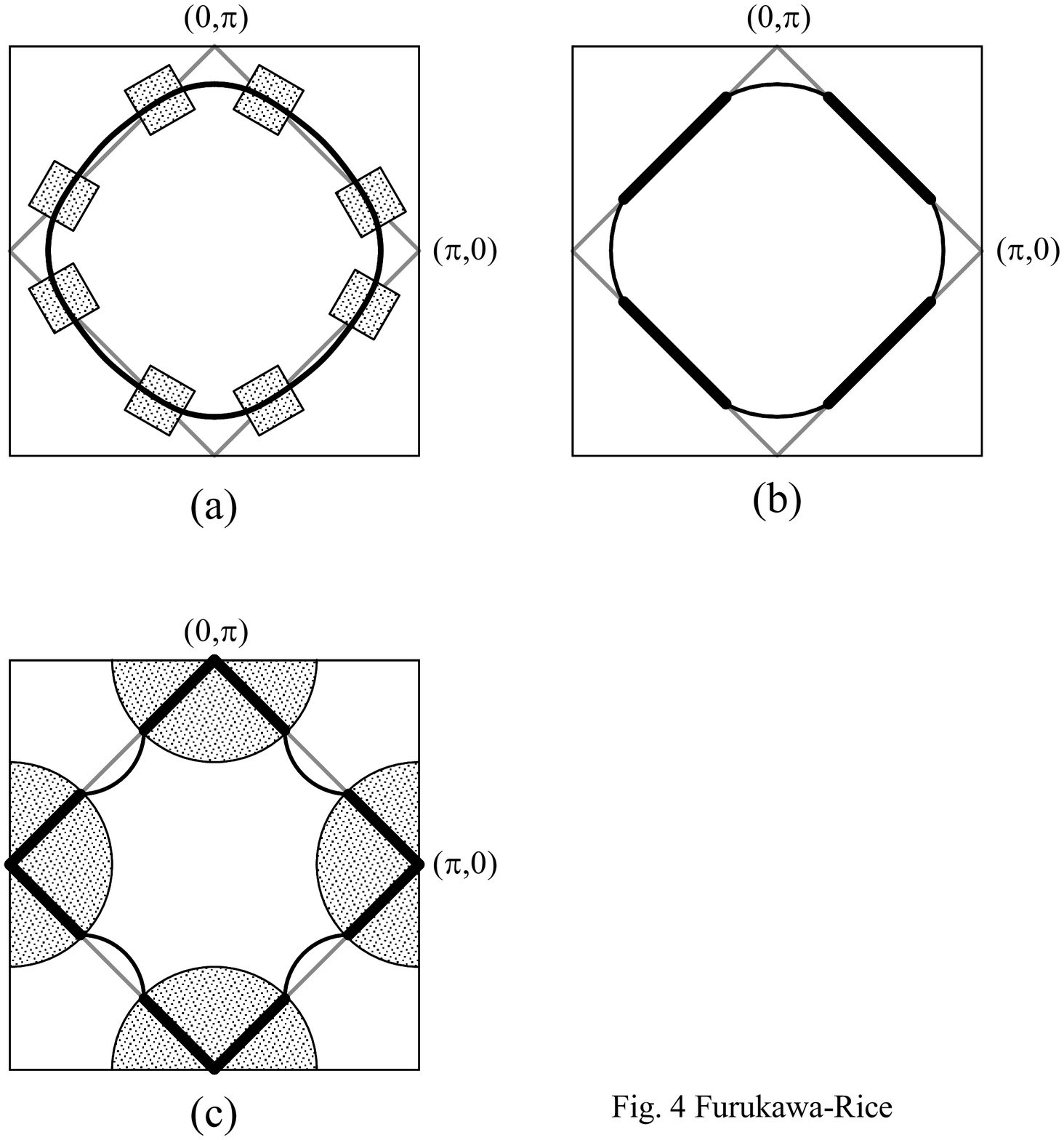}

Shapes of the Fermi surfaces when electron concentration is further
increased. (a) The noninteracting Fermi surface where 8 patches
are defined at the intersection with Umklapp surface.
(b) The distorted Fermi surface when it is pinned at 
$(\pm \pi/2, \pm\pi/2)$ points.
(c) In the case $t'<0$, the Fermi surface should be
pinned at $(\pi,0)$ and $(0,\pi)$.

\newpage
\epsfxsize=12cm
\epsfbox{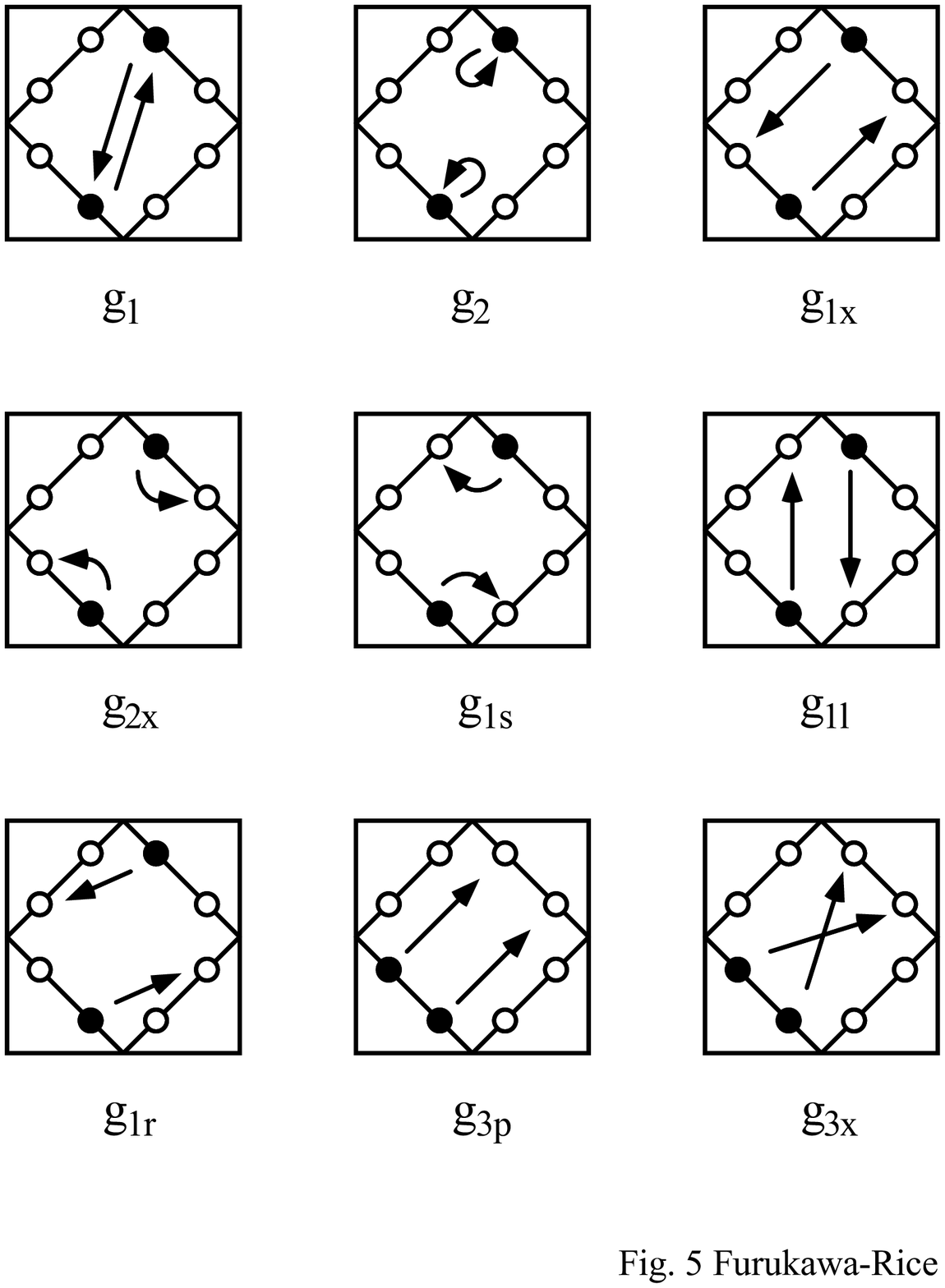}

The definitions of vertices for the 8-patch model.

\end{document}